\documentclass[showpacs,prl,aps,10pt,superscriptaddress,altaffilletter,floatfix,twocolumn]{revtex4-1}
\usepackage{graphicx}
\usepackage{hyperref}
\usepackage{amsmath}
\usepackage{amssymb}
\usepackage{color}
\usepackage{xspace}
\usepackage{dcolumn}
\usepackage{soul}
\usepackage{comment}
\newcolumntype{C}[1]{>{\centering\let\newline\\\arraybackslash\hspace{0pt}}m{#1}}
\usepackage[section]{placeins}

\RequirePackage{lineno}

\begin{document}

\newcommand{\isot}[2]{$^{#2}$#1}
\newcommand{\isotbold}[2]{$^{\boldsymbol{#2}}$#1}
\newcommand{\xeiso}{\isot{Xe}{136}\xspace}
\newcommand{\thsrc}{\isot{Th}{228}\xspace}
\newcommand{\cosrc}{\isot{Co}{60}\xspace}
\newcommand{\rasrc}{\isot{Ra}{226}\xspace}
\newcommand{\cssrc}{\isot{Cs}{137}\xspace}
\newcommand{\betascale}  {$\beta$-scale}
\newcommand{\kevkgyr}  {keV$^{-1}$ kg$^{-1}$ yr$^{-1}$}
\newcommand{\nonubb}  {$0\nu \beta\!\beta$\xspace}
\newcommand{\nonubbbf}  {$\boldsymbol{0\nu \beta\!\beta}$\xspace}
\newcommand{\twonubb} {$2\nu \beta\!\beta$\xspace}
\newcommand{\bb} {$\beta\!\beta$\xspace}
\newcommand{\vadc} {ADC$_\text{V}$}
\newcommand{\uadc} {ADC$_\text{U}$}
\newcommand{\mus} {\textmu{}s}
\newcommand{\chisq} {$\chi^2$}
\newcommand{\mum} {\textmu{}m}
\newcommand{\red}[1]{{\xspace\color{red}#1}}
\newcommand{\blue}[1]{{\xspace\color{blue}#1}}
\newcommand{\RunTwoA}{Run 2a}
\newcommand{\RunTwo}{Run 2}
\newcommand{\RunTwoBC}{Runs 2b and 2c}
\newcommand{\SP}[1]{\textsuperscript{#1}}
\newcommand{\SB}[1]{\textsubscript{#1}}
\newcommand{\SPSB}[2]{\rlap{\textsuperscript{#1}}\SB{#2}}
\newcommand{\pmasy}[3]{#1\SPSB{$+$#2}{$-$#3}}
\newcommand{\matel}{$M^{2\nu}$}
\newcommand{\psfac}{$G^{2\nu}$}
\newcommand{\tbeta}{T$_{1/2}^{0\nu\beta\beta}$}
\newcommand{\exolimit}[1][true]{\pmasy{2.6}{1.8}{2.1}$ \cdot 10^{25}$}
\newcommand{\exomeasurement}{\tbeta{}= \exolimit{}~yr}
\newcommand{\U}{\text{U}}
\newcommand{\V}{\text{V}}
\newcommand{\X}{\text{X}}
\newcommand{\Y}{\text{Y}}
\newcommand{\Z}{\text{Z}}
\newcommand{\bqcm}{${\rm Bq~m}^{-3}$}
\newcommand{\nonunorm}{N_{{\rm Err, } 0\nu\beta\beta}}
\newcommand{\nonunum}{n_{0\nu\beta\beta}}
\newcommand{\cussim}[1]{$\sim$#1}
\newcommand{\halflife}[1]{$#1\cdot10^{25}$~yr}
\newcommand{\numspec}[3]{$N_{^{#2}\mathrm{#1}}=#3$}
\newcommand{\TD}[1]{\textcolor{red}{#1}}
\newcommand{\PI}{Phase~I\xspace}
\newcommand{\PII}{Phase~II\xspace}
\newcommand{\Rn}{radon\xspace}
\newcommand\Tstrut{\rule{0pt}{2.6ex}} 

\setstcolor{blue}

\title{Measurement of the Spectral Shape of the $\beta$-decay of $^{137}$Xe to the Ground State of $^{137}$Cs in EXO-200 and Comparison with Theory}

\author{S.~Al Kharusi}\affiliation{Physics Department, McGill University, Montreal H3A 2T8, Quebec, Canada}
\author{G.~Anton}\affiliation{Erlangen Centre for Astroparticle Physics (ECAP), Friedrich-Alexander-University Erlangen-N\"urnberg, Erlangen 91058, Germany}
\author{I.~Badhrees}\altaffiliation{Permanent position with King Abdulaziz City for Science and Technology, Riyadh, Saudi Arabia}\affiliation{Physics Department, Carleton University, Ottawa, Ontario K1S 5B6, Canada}
\author{P.S.~Barbeau}\affiliation{Department of Physics, Duke University, and Triangle Universities Nuclear Laboratory (TUNL), Durham, North Carolina 27708, USA}
\author{D.~Beck}\affiliation{Physics Department, University of Illinois, Urbana-Champaign, Illinois 61801, USA}
\author{V.~Belov}\affiliation{Institute for Theoretical and Experimental Physics named by A.I. Alikhanov of National Research Centre ``Kurchatov Institute'', 117218, Moscow, Russia}
\author{T.~Bhatta}\affiliation{Department of Physics, University of South Dakota, Vermillion, South Dakota 57069, USA}
\author{M.~Breidenbach}\affiliation{SLAC National Accelerator Laboratory, Menlo Park, California 94025, USA}
\author{T.~Brunner}\affiliation{Physics Department, McGill University, Montreal H3A 2T8, Quebec, Canada}\affiliation{TRIUMF, Vancouver, British Columbia V6T 2A3, Canada}
\author{G.F.~Cao}\affiliation{Institute of High Energy Physics, Beijing 100049, China}
% ...


\date{\today}

\begin{abstract}
We report on a comparison between the theoretically predicted and experimentally measured spectra of the first-forbidden non-unique $\beta$-decay transition $^{137}\textrm{Xe}(7/2^-)\to\,^{137}\textrm{Cs}(7/2^+)$. The experimental data were acquired by the EXO-200 experiment during a deployment of an AmBe neutron source. The ultra-low background environment of EXO-200, together with dedicated source deployment and analysis procedures, allowed for collection of a pure sample of the decays, with an estimated signal-to-background ratio of more than 99-to-1 in the energy range from 1075 to 4175 keV. In addition to providing a rare and accurate measurement of the first-forbidden non-unique $\beta$-decay shape, this work constitutes a novel test of the calculated electron spectral shapes in the context of the reactor antineutrino anomaly and spectral bump.  
\end{abstract}


\maketitle

\paragraph{Introduction.} The discrepancies between measured and predicted antineutrino fluxes from nuclear reactors constitute the so-called reactor antineutrino anomaly~\cite{Mueller2011,An2017}. In addition, an event excess (``bump'') against predicted spectra between 4 and 7 MeV of antineutrino energy has been observed by the RENO~\cite{Ahn2012}, Double Chooz~\cite{Abe2014}, and Daya Bay~\cite{An2016} antineutrino-oscillation experiments. The spectral bump was apparently present, but not recognized then, in the much earlier Goesgen experiment~\cite{goesgen}. Predicting the reactor antineutrino flux is difficult due to the uncertainties related to the treatment of the $\beta$ decays of the numerous fission fragments~\cite{Hayes2014,Sonzogni2017}. One particular problem is the description of the forbidden $\beta$-decay transitions whose spectra are translated to antineutrino spectra at energies relevant for the measurement of the total flux and the spectral bump~\cite{Hayen2019a}. It has been noted that many first-forbidden $\beta$-decay transitions, like the presently discussed one, in the medium-mass $A=89-143$ nuclei play a key role in reactor antineutrino spectra~\cite{Hayen2019a,Hayen2019b}. Only a handful of electron spectra corresponding to $J^+\leftrightarrow J^-$ $\beta$ transitions in this region
has been measured and with a rather poor precision~\cite{Daniel1968,Booij1971}. 
According to~\cite{Hayen2019a,Hayen2019b} the $\beta$ spectra for the $J^+\leftrightarrow J^-$ transitions, relevant for solving the reactor anomaly and spectral bump, deviate noticeably from the allowed shape, the deviation being approximately a quadratic function of the electron kinetic energy (see, e.g., Ref.~\cite{Hayen2019b}, Figure 3, top panel). This is the case also for the $\beta$ decay of $^{137}$Xe (see Figure~\ref{fig:raw_vs_mcd}, lower panel), making this decay an important test case of the computed spectral shapes.
In the case of $^{137}$Xe there is a measurement~\cite{onega64} that proposes a scheme for the decay of $^{137}$Xe to the ground state and first excited state of $^{137}$Cs, but we could not find measurements or calculations of the corresponding $\beta$-spectrum shapes. In the present work we perform the $\beta$-spectrum-shape measurement and calculation for the decay to the ground state. Comparison with experiment confirms that the calculated shape of the $^{137}$Xe decay is correct, and thus there is hope that the effects of the first-forbidden $\beta$ decays lead to mitigation of the reactor anomaly and possible explanation of the origins of the spectral bump, as proposed by Hayen et al.~\cite{Hayen2019a,Hayen2019b}. 

The problem of many of the electron spectra of the first-forbidden $\beta$-decay transitions is connected to the uncertainty of the effective value of the weak axial coupling $g_{\rm A}$~\cite{Suhonen2017} and the enhancement of the axial-charge nuclear matrix element (NME) by meson-exchange currents~\cite{Kostensalo2018b}. Recently, a sustained effort has gone to clarifying these two burning issues~\cite{Ejiri2019}. Related to this, we point out that the effective values of $g_{\rm A}$ are more of effective corrections to specific nuclear-theory frameworks than fundamental corrections to the weak axial coupling~\cite{nature_phys}. For some decays the spectral shape depends on the effective value of $g_{\rm A}$ and, to some extent, on the mesonic enhancement~\cite{Suhonen2017,Kostensalo2018b,Ejiri2019}. The uncertainties related to these parameters are reflected as theoretical uncertainties in the predicted antineutrino spectra. Fortunately, the majority of the shapes of electron spectra are not much affected  by the values of these quantities. In order to test the accuracy of the theory framework used to compute the electron spectra related to the reactor-antineutrino problem one needs (a) a measured electron spectral shape of a forbidden $\beta$-decay transition in the nuclear mass region relevant for the reactor antineutrino problem with (b) a non-trivial shape and (c) independent of both $g_{\rm A}$ and the mesonic enhancement. 

The three requirements are met by the first-forbidden non-unique $\beta$-decay transition $^{137}\textrm{Xe}(7/2^-)\to\,^{137}\textrm{Cs}(7/2^+ GS)$. The condition
(a) is accounted for by the experimental spectral shape extracted in the present work. The condition (b) is satisfied by the complex spectral shape containing a pseudoscalar part with two NMEs, a pseudovector part with three NMEs and a pseudotensor part with one NME~\cite{Kostensalo2018b,Ejiri2019}. Furthermore, our present calculations, based on the formalism of~\cite{Behrens1982} and on its recent derivative~\cite{Haaranen2017}, show that point (c) is also satisfied to a high level of precision. 

\paragraph{Theoretical description of the forbidden $\beta$ shape.}For the theoretical description of the first-forbidden $\beta^-$ decay we adopt the expansion of Behrens and B{\"u}hring~\cite{Behrens1982}. NMEs up to next-to-leading order are included in the calculations~\cite{Haaranen2017}.

The nuclear-structure calculations were done using the shell model code NuShellX@MSU~\cite{nushellx} in a model space spanned by the proton orbitals $0g_{7/2}$, $1d_{5/2}$, $1d_{3/2}$, $2s_{1/2}$, and $0h_{11/2}$ and the neutron orbitals $0h_{9/2}$, $1f_{7/2}$, $1f_{5/2}$, $2p_{3/2}$, $2p_{1/2}$, and $0i_{13/2}$ with the effective Hamiltonian jj56pnb~\cite{jj56pnb}. This interaction has previously been used to study the mesonic-exchange effects on and $g_{\rm A}$-dependence of the electron spectra of $A\approx135$ nuclei~\cite{Kostensalo2018b}, as well as to predict the $\beta$ shapes of the first-forbidden decays contributing to the cumulative $\beta$ spectra from nuclear reactors~\cite{Hayen2019a,Hayen2019b}. While $^{137}$Xe is not one of the major contributors itself, the neighbouring nuclei, such as $^{136,137,138}$I and $^{139,140}$Xe, are~\cite{Chadwick2011}. The $^{137}$Xe ground-state-to-ground-state decay to $^{137}$Cs (GS decay) turns out to be one of the spectra with negligible shape dependence on the adopted value of $g_{\rm A}$ or the magnitude of the mesonic enhancement effects on the axial-charge matrix element.  
This is the case since the involved four axial-vector NMEs dominantly contribute to the spectral shape and thus $g_{\rm A}$ simply gives the overall scaling of the electron spectrum and, in turn, of the half-life. This $g_{\rm A}$ dominance is clearly visible in Figure~\ref{fig:raw_vs_mcd} where the $g_{\rm A}$ dependent contribution (blue dots) is compared with the full spectral shape (blue dotted line). The shape factor $C(E)$ (ratio of the corrected spectrum to that corresponding to an allowed decay) is plotted in the bottom frame of the figure.
This transition is a perfect test case for the accuracy and validity of the calculations of the $\beta$ spectra in the context of the reactor antineutrino anomaly~\cite{Hayen2019a,Hayen2019b}. This is particularly important since the calculations of Hayen et al.~\cite{Hayen2019a,Hayen2019b} propose corrections to the traditional Huber-Mueller model~\cite{hb1,Mueller2011} which explain, at least partially, the anomaly and spectral bump.

In contrast with the GS decay, the spectral shape of the $^{137}$Xe decay to the first excited state of $^{137}$Cs (ES decay) does depend on the value of $g_{\rm A}$ and could, in principle, be used to constrain its value. However, the accompanying emission of a de-excitation $\gamma$ makes accurate measurement of the ES decay’s $\beta$-spectrum shape in EXO-200 challenging. Since both the motivation and analysis approach are substantially different for the GS and ES measurements, we consider the ES decay outside the scope of this work and only focus on the GS decay.

\paragraph{Experimental details and results.}The EXO-200 detector is a cylindrical time projection chamber (TPC). It is filled with liquid xenon (LXe), consisting of 80.6\% of the isotope $^{136}$Xe and 19.1\% of $^{134}$Xe, with the remaining balance comprised of other isotopes. The LXe is housed in a cylindrical copper vessel of $\sim$40 cm diameter and $\sim$44 cm length. The vessel is surrounded by $\sim$50 cm of HFE~\cite{m3m}, a hydrogen-rich heat transfer fluid maintained inside a vacuum-insulated copper cryostat. Further shielding is provided by at least 25 cm of lead in all directions. A small diameter copper tube runs from the outside of the lead shield through the HFE and wraps around the outside of the TPC vessel. It allows one to insert miniature radioactive calibration sources and place them close to the active volume of the detector. Energy depositions in the TPC produce ionization charge and scintillation light. The charge and light signals are reconstructed to provide energy and position of events. In a given event, charge deposits, or clusters, that are separated by $\sim$1 cm can be individually reconstructed. The event is then classified as single-site (SS), or multi-site (MS), depending on the number of spatially distinct reconstructed clusters. More details about the EXO-200 design and performance are available in~\cite{Auger:2012gs,jinst2}. The reconstruction, Monte Carlo (MC) simulation, and analysis approaches are described in~\cite{Albert2013,exo_final}. EXO-200 is designed to minimize radioactive backgrounds. Its data rate above 1000 keV is dominated by the two-neutrino double $\beta$ decay of $^{136}$Xe~\cite{Ackerman:2011gz}. 

\begin{figure}[htbp]
    \centering
    \includegraphics[width=0.45\textwidth]{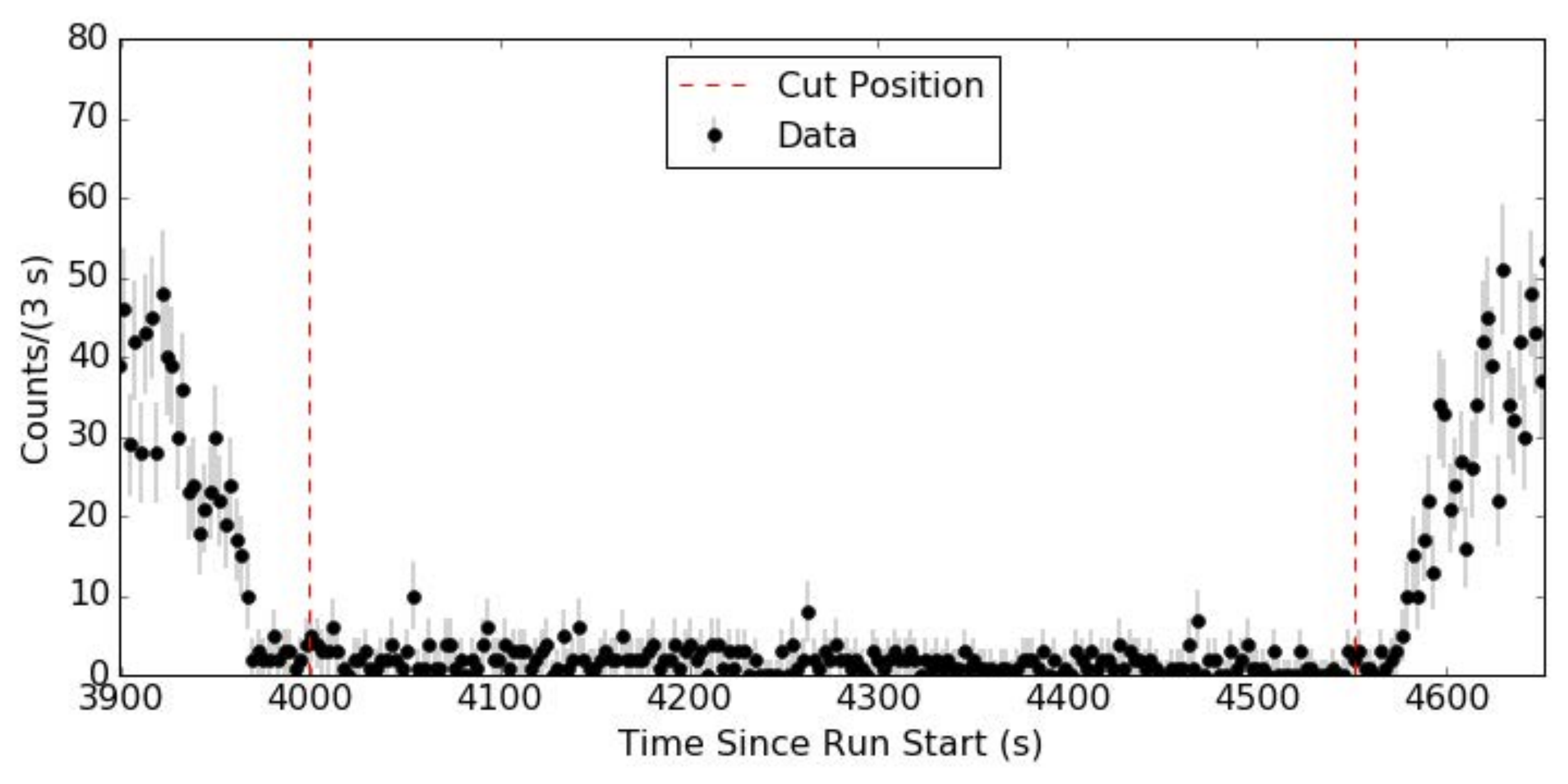}
    \caption{Reconstructed event rate during the AmBe source calibration. The vertical lines show the cuts that select $^{137}$Xe decays.}
    \label{fig:rate}
\end{figure}

\begin{figure}[htbp]
    \centering
    \includegraphics[width=0.45\textwidth]{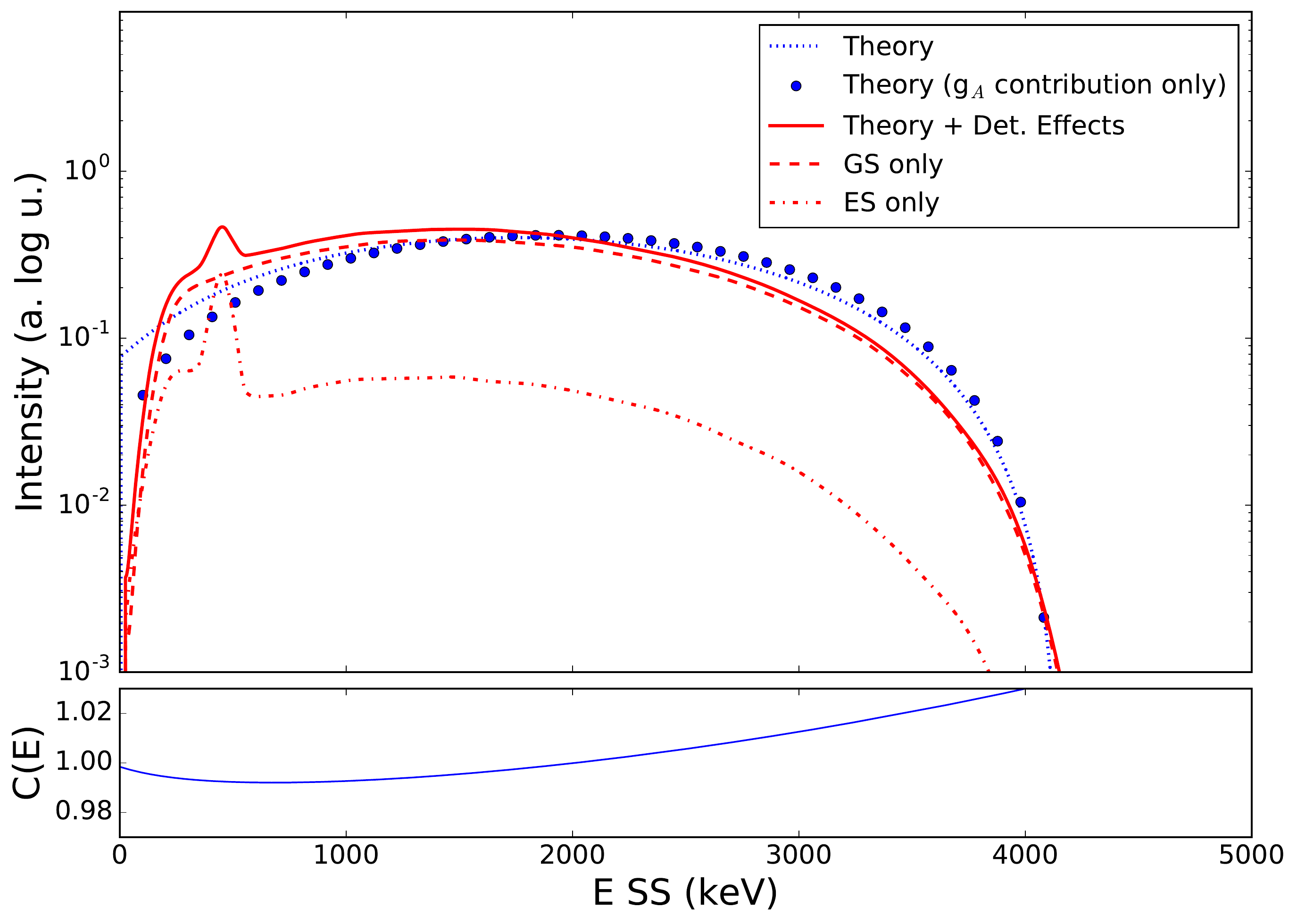}
    \caption{(Top frame) Theoretical GS spectrum (blue dotted line) and reconstructed MC SS spectrum (red solid line). The theoretical GS spectrum shape is the same for all reasonable $g_{\rm A}$ and mesonic enhancements within the line width. The  $g_{\rm A}$-dependent contribution to the theoretical spectrum is also shown as blue dots. Individual contributions of GS and ES decays to the reconstructed spectrum are also shown as red dashed and dash-doted lines, respectively. (Bottom frame) Shape factor, $C(E)$.}
    \label{fig:raw_vs_mcd}
\end{figure}

The experimental data used in this work were collected during the AmBe neutron source calibration campaign carried out in December 2018. $^{137}$Xe is produced by neutron capture on $^{136}$Xe and decays to $^{137}$Cs with the half-life of 3.818$\pm$0.013 min~\cite{t12}. In $\sim$67\% of cases~\cite{rasco}, $^{137}$Xe decays to the ground state of $^{137}$Cs. In $\sim$31\% of cases, $^{137}$Xe decays to a 5/2$^+$ excited state of $^{137}$Cs, which de-excites by emission of a 455.5 keV $\gamma$-ray. 
The neutrons were produced by the neutron source positioned at the mid-plane of the TPC, 3 cm outside the LXe volume. The source contains $\sim$65 $\mu$Ci of $^{241}$Am in the form of a carrier-free $^{241}$AmO$_2$ powder mixed with beryllium metal powder. The mixture is contained in a 1.2 mm diameter tungsten capsule, which is in turn contained inside a 2.0 mm diameter stainless steel capsule that is welded shut by electron-beam welding. The estimated neutron activity of the source is $\sim$90 Bq. 
More details about the source construction and characterization can be found in~\cite{disser}. In $\sim$60\% of the cases~\cite{ratio}, the neutron emission from the source is accompanied by a 4439.8 keV $\gamma$-ray. The source is positioned several centimeters outside of the TPC during the calibrations, which leads to some neutrons being captured in HFE by hydrogen nuclei. The capture is followed by the emission of a 2224.6 keV $\gamma$-ray. Additional $\gamma$ radioactivity is expected from neutron inelastic scattering in HFE. While advantageous for the energy calibration, the $\gamma$-rays produced when the AmBe source is deployed close to the TPC would constitute a major background for the $^{137}$Xe $\beta$ decay measurement. To avoid this, a special deployment procedure was used. The deployment sequence consisted of repeated ``$^{136}$Xe activation --- $^{137}$Xe decay'' cycles. During the decay phases, the source was retracted outside of the lead shield, ceasing the associated $\gamma$ radioactivity. The length of the periods was chosen to maximize the number of detected $^{137}$Xe decays. Figure~\ref{fig:rate} shows the rate of reconstructed events in EXO-200 during one of the decay periods when the source is retracted. The drop (rise) of the rate at the end (beginning) of the activation periods is clearly seen. The red lines indicate the placement of the cuts to select the $^{137}$Xe decay period. A total of 60 such periods is selected during the campaign. The decay phase is defined as a period when the event rate is less than 1.33 Hz. The timing cuts are placed at +30 (-30) seconds from each decay period's start (end). The integrated livetime is 8.73 hours.

The fiducial volume cuts are relaxed slightly, as compared to Ref.~\cite{exo_final}. This increases the fiducial mass by $\sim$5\%, while still retaining the good agreement between shapes of energy distributions in data and MC. The relaxed cuts admit a background increase that is negligible for this study.

When $^{137}$Xe decays to the ground state of $^{137}$Cs, only the $\beta$ particle is emitted and detected. Electrons of $O$(MeV) energy are reconstructed predominantly as SS events in the detector. On the other hand, when the decay proceeds to the 5/2$^+$ excited state (ES decay), both the $\beta$ and the de-excitation $\gamma$ deposit energy and are reconstructed as an MS event in most cases. Therefore, the $^{137}$Xe GS decay spectrum can be examined in EXO-200 by looking at the energy distribution of the selected SS events. However, several reconstruction and physics related effects introduce non-negligible differences between the theoretical GS spectrum and the spectrum of the reconstructed SS events. To take these effects into account, the MC of the AmBe source is first used to track the neutrons up to the $^{136}$Xe atoms on which they are captured. $^{137}$Xe decays, both GS and ES, are then generated from the capture position distributions. The $\beta$ energy is sampled from the theoretical $\beta$ spectrum.
The decay products ($\beta$ and de-excitation $\gamma$) are tracked, and their energy depositions are simulated and reconstructed to produce the expected SS spectrum. This spectrum, along with the theoretical one, are shown in Figure~\ref{fig:raw_vs_mcd}. 
At the lowest energy one can see the expected effect of the charge reconstruction threshold, leading to the MC spectrum having a lower intensity than the theoretical spectrum. While the SS spectrum is dominated by the GS decays, a residual peak at 455.5 keV is expected, due to ES decays that occur outside of the sensitive volume. For such events, the $\beta$ cluster of an ES decay is lost, while the de-excitation $\gamma$-ray has a chance to travel to the fiducial region and get reconstructed as a single cluster. At higher energies, the intensity of the MC SS spectrum is lower than the theoretical spectrum, due to reconstruction related effects and the production of Bremsstrahlung photons by the $\beta$ particles, which leads to some GS decays being reconstructed as MS events. Finally, the slightly higher apparent end-point in the MC spectrum is expected, due to the finite energy resolution.

\begin{figure}[htbp]
    \centering
    \includegraphics[width=0.45\textwidth]{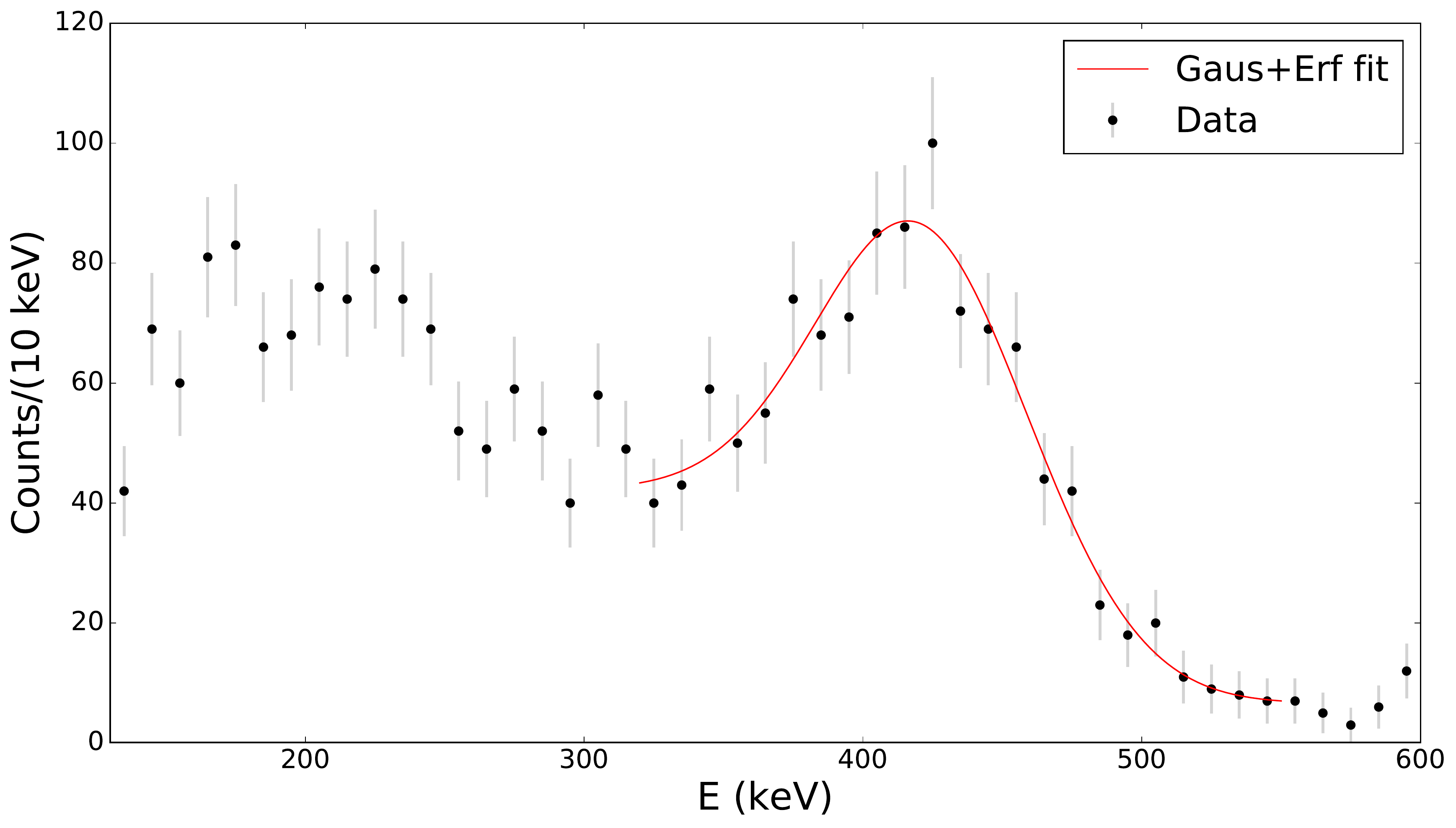}
    \caption{Selected $^{137}$Cs de-excitation $\gamma$ events. The Gaussian+Erf fit to the uncalibrated charge energy is shown as the red line.}
    \label{fig:455uncalib}
\end{figure}

\begin{figure}[htbp]
    \centering
    \includegraphics[width=0.45\textwidth]{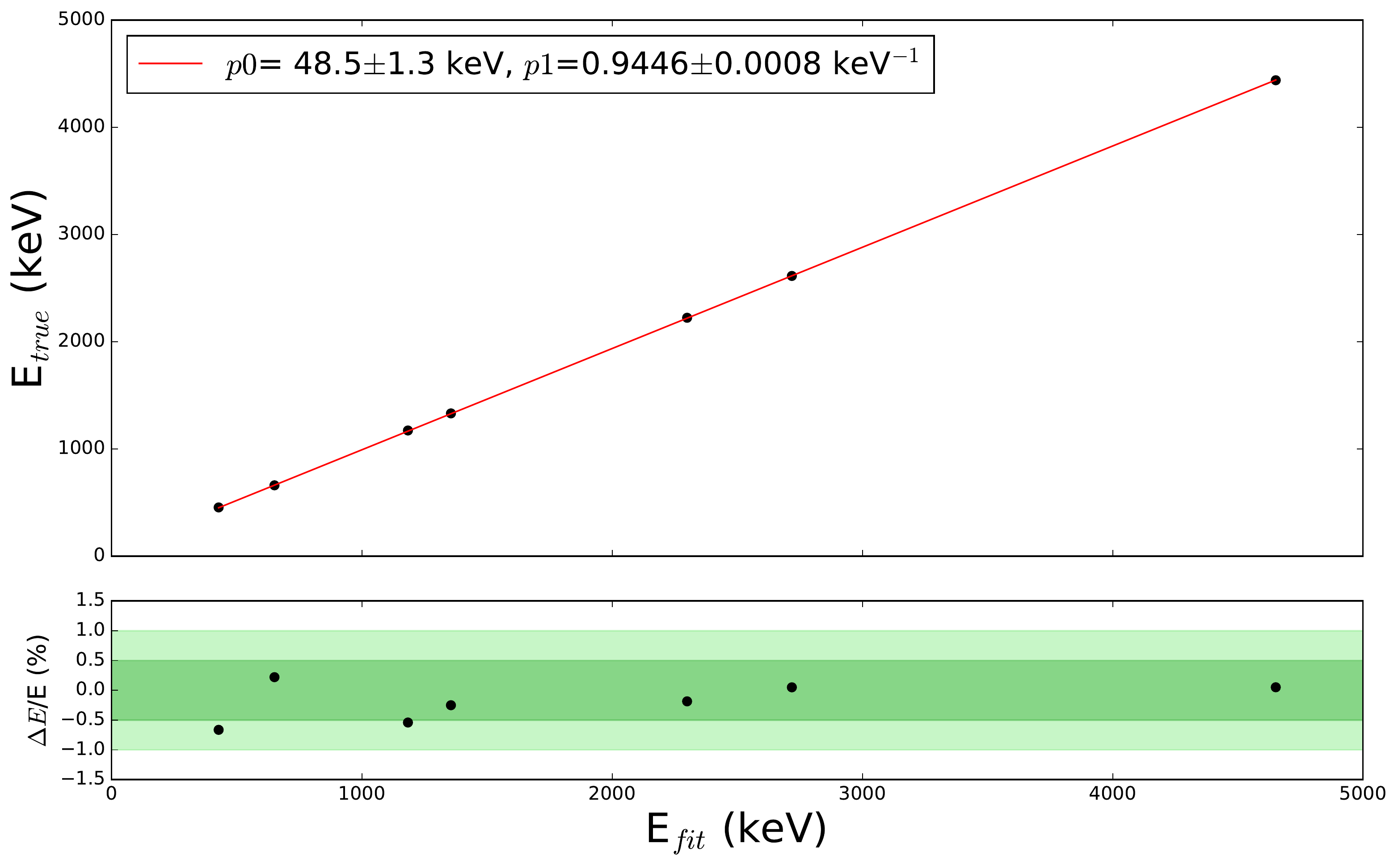}
    \caption{SS data energy calibration. Red line is the linear fit. Best-fit parameters are also shown. The errors are statistical.}
    \label{fig:ss_calib}
\end{figure}

The detector's energy scale is constrained using the total of seven mono-energetic $\gamma$ lines obtained in EXO-200 using radioactive calibration sources: 455.5 (AmBe), 661.7 ($^{137}$Cs), 1173.2 ($^{60}$Co), 1332.5 ($^{60}$Co), 2224.6 (AmBe), 2614.5 ($^{228}$Th), and 4439.8 keV (AmBe). The mean position of the full absorption peaks in the uncalibrated energy spectra is found using a fit by a linear combination of the Gaussian and error functions. The latter function is an ad-hoc way to account for the shoulder to the left of the peaks, comprised of Compton scattering events, multi-site full absorption events with one or more small charge clusters missing, and other events. The calibration runs collected closest in time to the AmBe calibration campaign are used. The same fiducial cuts are used for the calibration events as for the $^{137}$Xe dataset. The SS events are selected for all calibration lines, with the exception of the 455.5 keV $^{137}$Cs de-excitation line. Since in that case the de-excitation $\gamma$ is accompanied by a $\beta$ decay, the two-cluster MS events within the timing cuts are first selected. The energy distribution of the smaller of the two charge clusters is then plotted for events in which the larger of the two charge clusters has energy $\sim$3$\sigma$ above 455.5 keV (560 keV). Figure~\ref{fig:455uncalib} shows the resulting spectrum. 
It is not possible to discern contributions of individual clusters to the total detected scintillation light. So the reconstructed energy in this work is based on charge signals only. 
The energy calibration approach used in this work extends the constrained energy range in both directions, as compared to previous analyses, at the expense of a worse energy resolution. 
After the mean positions of all $\gamma$ lines are found, they are plotted versus the true energies and fit by a linear function. Figure~\ref{fig:ss_calib} shows the resulting SS data energy calibration that is used in this analysis.
The residuals are typically within $\pm$0.5\%, not exceeding $\pm$1\%.

\begin{figure}[htbp]
    \centering
    \includegraphics[width=0.45\textwidth]{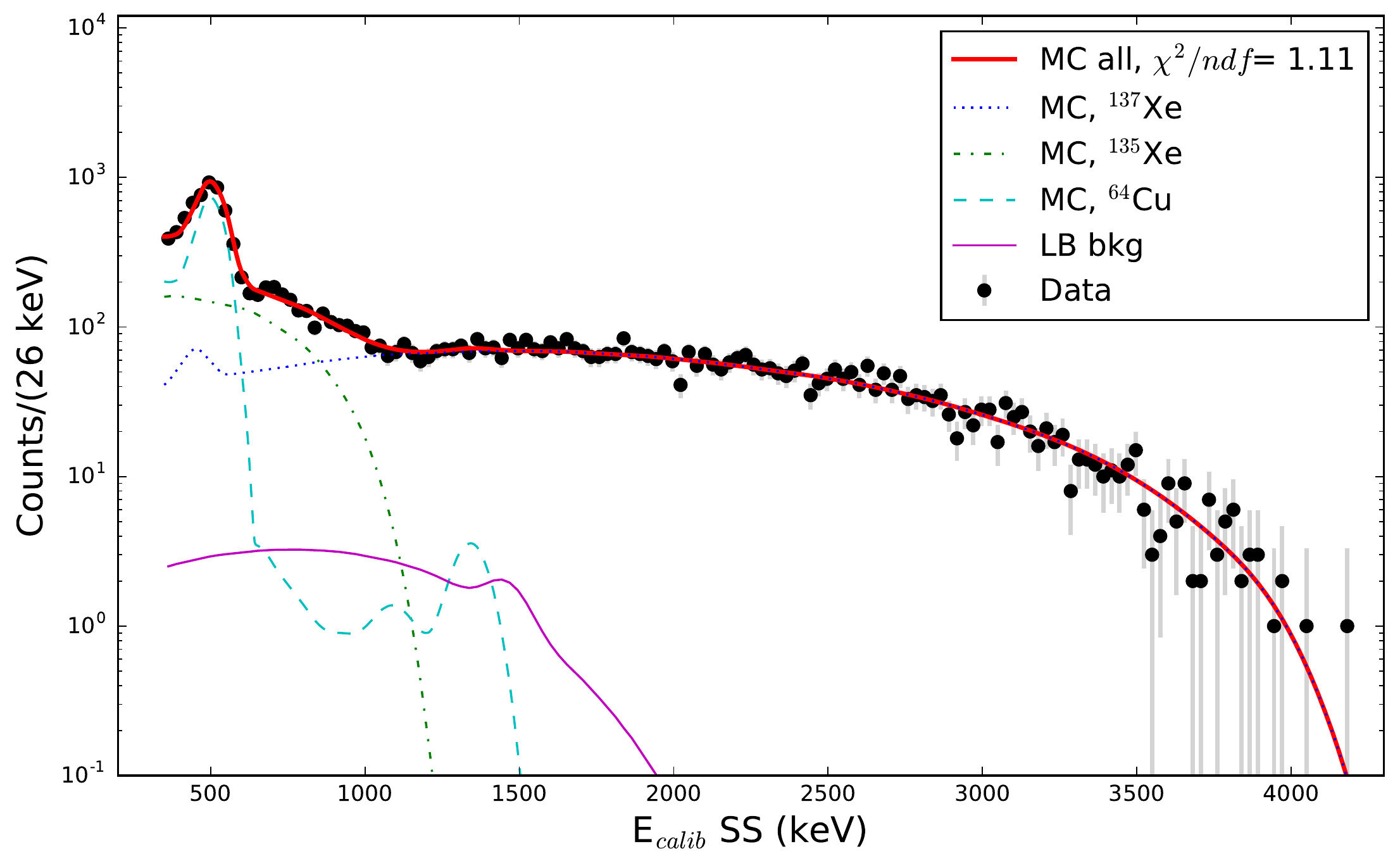}
    \caption{Calibrated SS energy spectrum of events passing the selection cuts (black points). Blue dotted, green dashed, and cyan dot-dashed lines correspond to MC spectra of $^{137}$Xe, $^{135}$Xe, and $^{64}$Cu, respectively. Thick red line corresponds to the sum of the three best-fit components. Thin magenta line corresponds to the LB backgrounds, described in the text. The reduced $\chi^2$ of the fit is shown in the legend.}
    \label{fig:3fit}
\end{figure}

Based on MC of the AmBe source, the main expected backgrounds in the selected SS spectrum are $^{135}$Xe and $^{64}$Cu. $^{135}$Xe is produced by capture of the AmBe neutrons on $^{134}$Xe, which constitutes $\sim$19\% of the xenon target in EXO-200. $^{135}$Xe  undergoes a $\beta$ decay with a half-life of 9.14 hours and has a Q-value of 1051 keV. $^{64}$Cu is produced in the copper vessel (and other construction elements) and undergoes a $\beta$+/EC decay with a half-life of 12.7 hours. Only a single 511 keV positron annihilation $\gamma$-ray is expected to be seen in the SS spectrum. In $\sim$0.5\% of cases, $^{64}$Cu electron captures to an excited state of $^{64}$Ni that de-excites by a 1345.8 keV $\gamma$-ray, which can also produce an SS event. The expected SS spectra of $^{135}$Xe and $^{64}$Cu are generated by MC analogously to the case of $^{137}$Xe. The three spectral shapes are then fit to the calibrated charge energy spectrum of the selected SS events allowing the normalization of each of the three components to float. Figure~\ref{fig:3fit} shows the selected SS events and the results of the fit. 
The good agreement between the best-fit and the data shapes supports the expectation that $^{64}$Cu and $^{135}$Xe are the main activation backgrounds. 

\begin{figure}[htbp]
    \centering
    \includegraphics[width=0.45\textwidth]{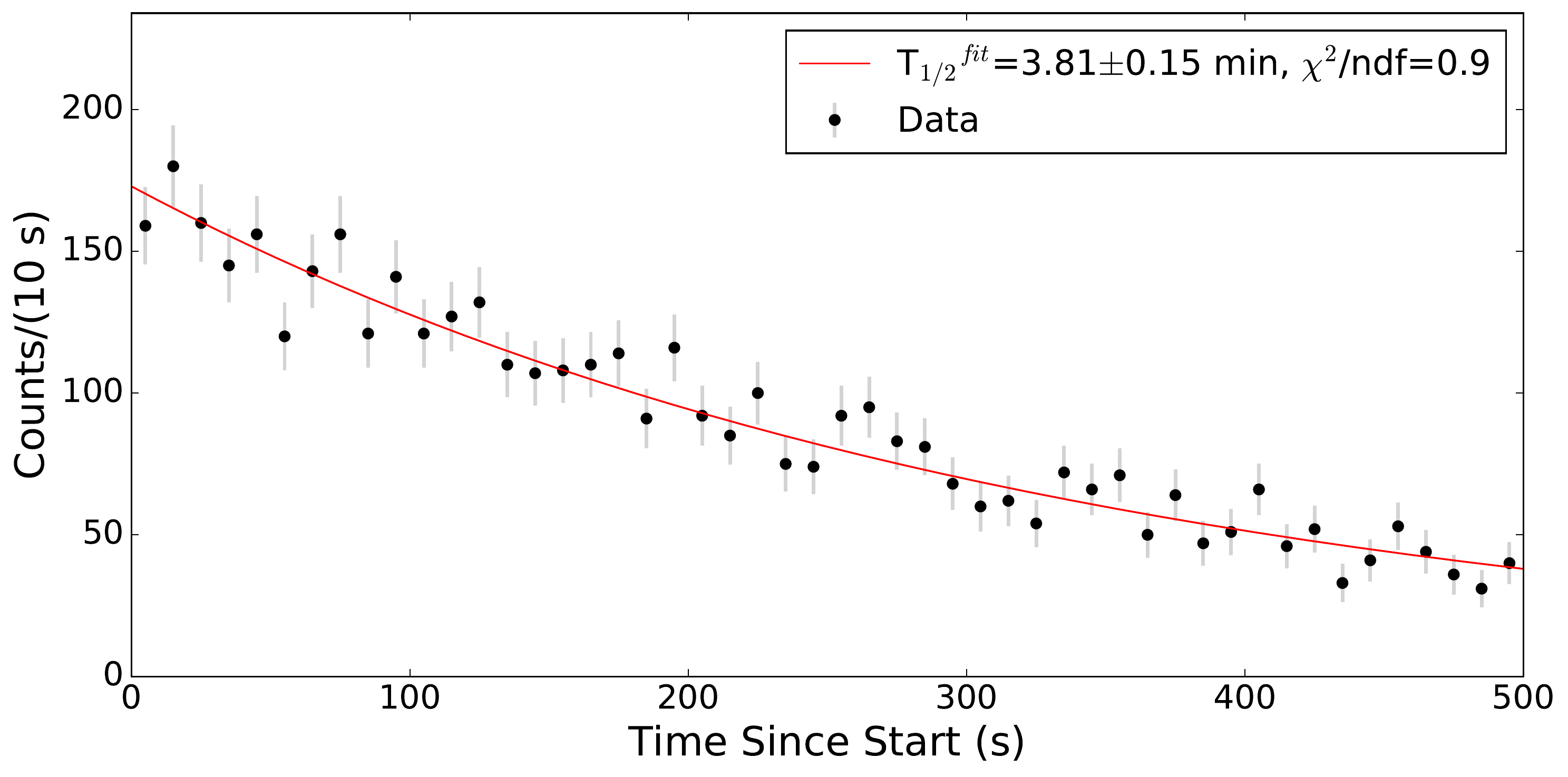}
    \caption{Time distribution of selected SS events (black) with energies between 1075 and 4175 keV. The exponential fit is shown as red solid line. 
    }
    \label{fig:dt}
\end{figure}

A SS low energy cut of 1075 keV is chosen to remove the $^{135}$Xe and most of the $^{64}$Cu events. The high energy cut is set to 4175 keV, based on the Q-value of $^{137}$Xe GS decay. Based on the fit, the residual background contribution of $^{64}$Cu and $^{135}$Xe to the selected energy range is 22.7(5) and 0.50(2) events, respectively. Two known background contributions to the AmBe dataset are two-neutrino double $\beta$ and $^{40}$K decays, whose rates are constrained by the EXO-200 ``low background data'' (LB)~\cite{exo_final}. Taking into account the livetime and the correction for the slightly larger fiducial volume used in this analysis, one expects 43 two-neutrino double $\beta$ and 7.8 $^{40}$K events, or $\sim$1.1\% of all the SS events in the selected energy range. Other LB components are expected to contribute less than 1 event total. The rate of the LB events is known to $\sim$10\% relative uncertainty. The expected LB events are removed from the dataset by subtracting their MC spectra, normalized to the corresponding number of expected events. The remaining dataset contains 4526 events.
For a qualitative check of the purity of the selected dataset, the time difference between each selected event and the start time of the corresponding decay period is histogrammed and fit by an exponential function (Figure~\ref{fig:dt}).  
The best-fit half-life value, 3.81$\pm$0.15 min, is in good agreement with the known half-life of $^{137}$Xe of 3.818$\pm$0.013 min~\cite{t12}. 

\begin{figure*}[htbp]
    \centering
    \includegraphics[width=0.75\textwidth]{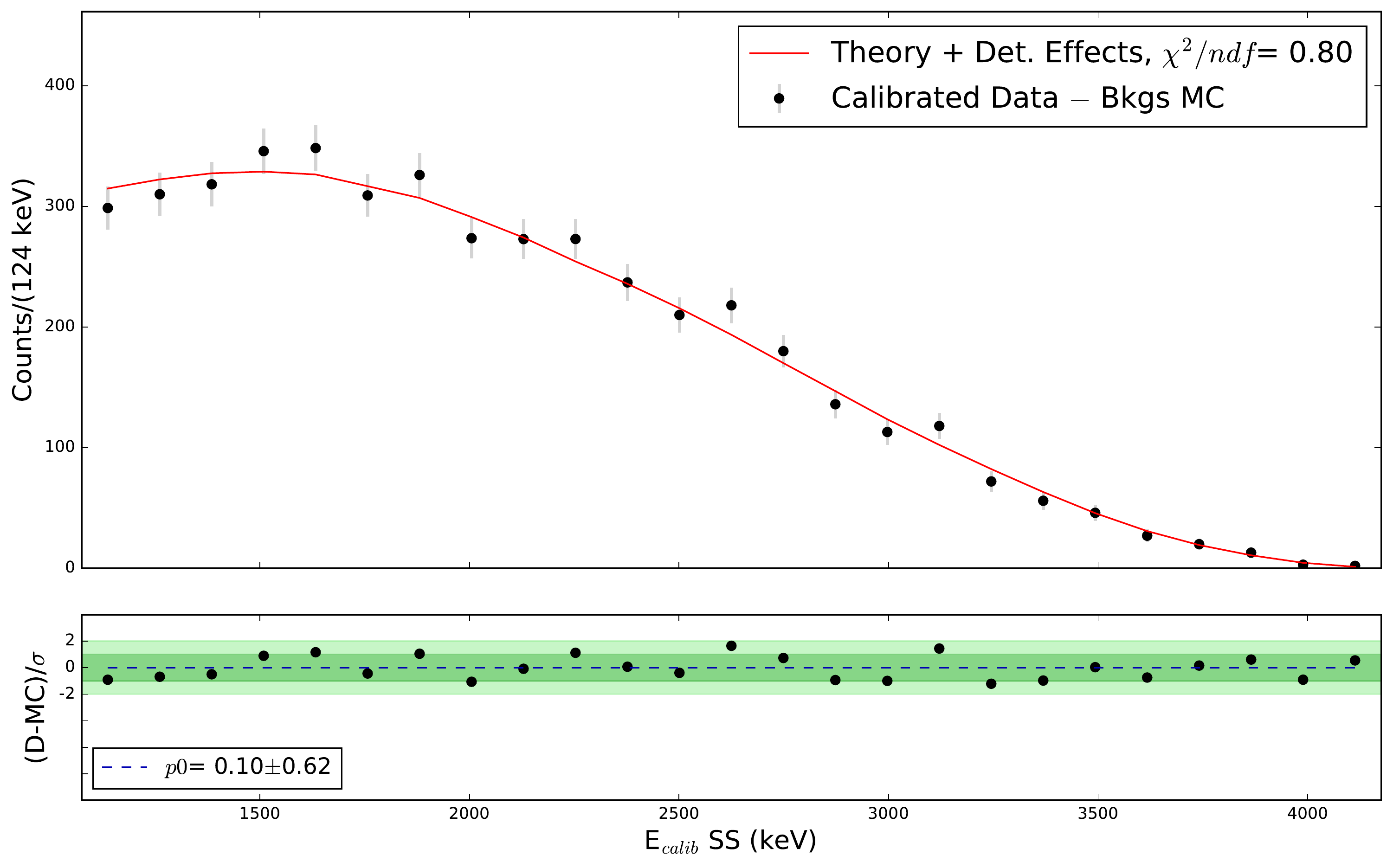}
    \caption{(Top frame) Best fit to the selected, calibrated, background-subtracted SS data events. The data points are shown in black. The theoretical spectrum (after passing through MC) is shown in red. (Bottom frame) Residual differences between the data and best-fit curve, normalized by the statistical errors, are shown in black. The constant 
    fit to the residuals is shown by dashed blue line. $p0$ corresponds to the constant term of the fit.}
    \label{fig:results}
\end{figure*}

Figure~\ref{fig:results} shows the comparison between the observed and expected GS spectra of the $^{137}$Xe events. The comparison range is from 1075 to 4175 keV. 
The calibrated charge energy spectrum of the selected SS data events, with the expected residual background contributions subtracted, is shown in black on the top frame of the figure. It is fit with the simulated shape based on the theoretical calculation (red). The only parameter floating in the fit is the total normalization. The reduced $\chi^2$ of the fit (also shown) suggests a good agreement between the data and expectation. The normalized residuals are shown on the bottom frame of the figure. All residuals are within $\pm$2 $\sigma$ statistical error. The residuals are fit by a constant (dashed blue line) 
trend line, with the best-fit parameter shown. The residuals show no  statistically significant energy dependence.

Anything that can introduce an energy-dependent discrepancy between the data and MC can systematically affect the comparison shown on Figure~\ref{fig:results}. Given the amount of the available statistics, we are sensitive to potential systematics effects on the level of a few percent or more. The data energy calibration is constrained to the sub-percent level. The Gaussian+Erf fit model itself may be a source of systematics when extracting the peaks mean positions. This effect was studied by EXO-200 and is expected to introduce a $\sim$3 keV bias, which is sub-dominant to the calibration residuals (Figure~\ref{fig:ss_calib}). The residual background contamination in the selected SS energy range contributes $\leq$1\% of events and is known to $O$(10\%) relative uncertainty, suggesting only a fraction of percent residual effect. Potential imperfections of the MC and reconstruction can systematically affect the comparison only if they lead to an energy-dependent difference of the SS fraction or of the overall SS spectral shape in the data and MC. Based on the latest published comparison of data and MC in EXO-200 (Figure 1 in Ref.~\cite{exo_final}), the energy-dependent deviation is expected to be small, compared to the statistical errors in Figure~\ref{fig:results}.

\paragraph{Discussion and conclusion.}
We calculate the $^{137}$Xe GS spectrum and find that it has no significant dependence on the adopted value of $g_{\rm A}$ or the magnitude of the mesonic enhancement effects on the axial-charge matrix element. This makes this transition an ideal tool to validate the accuracy of the $\beta$ spectra calculations in the context of the reactor antineutrino anomaly. We perform a precise measurement of this first forbidden non-unique $\beta$-decay shape using the data collected during an AmBe source deployment in EXO-200. A good agreement between the predicted and observed spectra is found. Therefore, this work provides both a rare measurement of the first forbidden non-unique $\beta$-decay shape and a novel test related to the calculated electron spectral shapes of beta decays that contribute strongly to the antineutrino flux from nuclear reactors. The hope is that this test justifies the calculated spectral shapes of~\cite{Hayen2019a,Hayen2019b} thus implying that the spectral bump and the flux anomaly could be explained, at least partly, by the exact spectral shapes of the abundant first-forbidden non-unique beta decays of the fission fragments in nuclear reactors.

\paragraph{Acknowledgments.}
EXO-200 is supported by DOE and NSF in the United States, NSERC in Canada, IBS in Korea, RFBR (18-02-00550) in Russia, DFG in Germany, and CAS and ISTCP in China. EXO-200 data analysis and simulation uses resources of the National Energy Research Scientific Computing Center (NERSC). This work has been partially supported by the Academy of Finland under the Academy project no. 318043. J. K. acknowledges the financial support from the Jenny and Antti Wihuri Foundation. We gratefully acknowledge the KARMEN collaboration for supplying the cosmic-ray veto detectors, and the WIPP for their hospitality. 

\bibliography{Xe137_GS_paper}

\end{document}